\documentclass[apl,twocolumn,floatfix,superscriptaddress,citeautoscript]{revtex4}

\usepackage{graphicx}
\usepackage{dcolumn}   
\usepackage{bm}        
\usepackage{amssymb}   
\usepackage{verbatim}
\usepackage{multirow}

\usepackage{amsmath}
\usepackage{amsfonts}

\begin{document}

\title{Reducing microwave loss in superconducting resonators due to trapped vortices}

\author{C. Song}
\affiliation{Department of Physics, Syracuse University, Syracuse, NY 13244-1130}
\author{M.P. DeFeo}
\affiliation{Department of Physics, Syracuse University, Syracuse, NY 13244-1130}
\author{K. Yu}
\affiliation{Department of Physics, Syracuse University, Syracuse, NY 13244-1130}
\author{B.L.T. Plourde}
\email[]{bplourde@phy.syr.edu}
\affiliation{Department of Physics, Syracuse University, Syracuse, NY 13244-1130}

\date{\today}

\begin{abstract}
Microwave resonators with high quality factors have enabled many recent breakthroughs with superconducting qubits and photon detectors, typically operated in shielded environments to reduce the ambient magnetic field. 
Insufficient shielding or pulsed control fields can introduce vortices, leading to reduced quality factors, although increased pinning can mitigate this effect. 
A narrow slot etched into the resonator surface 
provides a straightforward method for pinning enhancement without otherwise affecting the resonator. Resonators patterned with such a slot 
exhibited over an order of magnitude reduction in the excess loss due to vortices compared with identical resonators from the same film with no slot.
\end{abstract}

\maketitle

Low-loss microwave resonators fabricated from superconducting thin films are playing key roles in many recent low-temperature experiments. 
Such resonators can be coupled to quantum coherent superconducting devices, or qubits \cite{clarke08}, for explorations of QED with circuits \cite{wallraff04, hofheinz09}. 
Furthermore, high-quality factor resonators have enabled the development of Microwave Kinetic Inductance Detectors (MKIDs), highly sensitive photon detectors for astrophysical measurements \cite{zmuidzinas03}. 

A variety of factors determine the quality factor of these resonators, including dielectric loss in the substrates and thin-film surfaces \cite{oconnell08}. If the resonators are not cooled in a sufficiently small ambient magnetic field, or if large pulsed fields are present for operating circuits in the vicinity of the resonators, vortices can become trapped in the resonator traces, thus providing another loss channel. The presence of even a few vortices can substantially reduce the resonator quality factor \cite{song09}. 

Vortex dynamics at microwave frequencies has been studied for some time both theoretically \cite{coffey91, brandt91} and experimentally in a variety of superconductors \cite{gittleman68, pompeo08, zaitsev07, belk96, song09}. 
The response of vortices to an oscillatory Lorentz force is determined primarily by two forces: the viscous force, due to the motion of the vortex core and characterized by a vortex viscosity $\eta$; 
and the pinning forces in the material that impede the vortex motion and, in the simplest case, can be described by a linear spring constant $k_p$. 
The ratio of the pinning strength to the vortex viscosity, $f_d = k_p/2\pi \eta$, determines the crossover frequency separating elastic and viscous response of the vortices. 

In this letter we demonstrate a technique for patterned pinning on the resonator surface to reduce the excess microwave loss due to trapped vortices. We compare field-cooled measurements for a series of coplanar waveguide (CPW) resonators patterned from the same thin film of Al, a common material used for qubits and MKIDs. Some of the resonators had a single longitudinal slot partially etched into the surface along the center conductor, while others had no slot. 
A surface step in a superconductor can pin a vortex because of the line energy variation associated with the change in thickness \cite{daldini74, plourde02}.  
Orienting the slot, with its surface steps on either side, along the length of the center conductor of a CPW resonator, aligns the pinning forces from the slot to oppose the Lorentz force due to the microwave current that flows in the resonator. 
Intuitively, one would expect such a slot to be most effective with a width comparable to the vortex core size, determined by the coherence length, which is of the order of a few hundred nm in these films \cite{song09}. 

\begin{figure}
\centering
\includegraphics[width=3.35in]{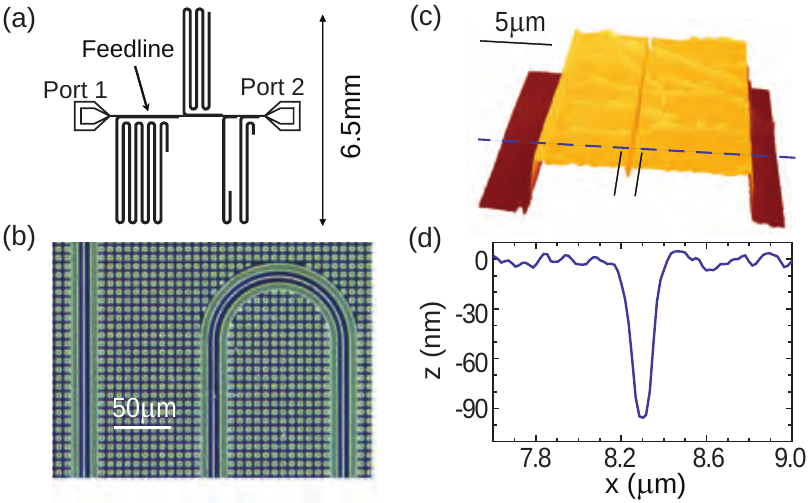}
  \caption{(Color online) (a) Chip layout showing common feedline and four resonators. (b) Darkfield optical micrograph near shorted end of resonator. (c) Atomic Force Microscope (AFM) image showing configuration of slot on center conductor. (d) AFM line trace across slot.
\label{fig:setup}}
\end{figure}

The resonator layout is identical to that in our previous work \cite{song09}, with four quarter-wave CPW resonators of lengths corresponding to fundamental resonances near 1.8, 3.3, 6.9, and 11.0 GHz, as calculated with the Sonnet microwave circuit simulation software \cite{sonnet05}. These resonators are coupled capacitively to a common CPW feedline, thus allowing for frequency multiplexing [Fig. \ref{fig:setup}(a)] \cite{mazin04,zmuidzinas03}, 
although we will focus on the resonators near 1.8 GHz in this work. 
We design for the resonators to be somewhat over-coupled at $B=$ 0, where the intrinsic loss at the measurement temperature (310 mK) is dominated by thermal quasiparticles. This gives us the ability to continue to resolve the resonances with the anticipated enhanced levels of loss once vortices are introduced. 

A 150 nm-thick Al film was electron-beam evaporated onto on a 2-inch sapphire wafer for patterning into multiple resonator chips with the layout of Fig. \ref{fig:setup}(a, b). 
On three of the dies, slots with a width of $200$~nm, along with alignment marks, were patterned using electron-beam lithography and reactive ion etching (RIE) in a combination of BCl$_3$, Cl$_2$, and CH$_4$, with the etch timed to stop at a depth of 90 nm [Fig. \ref{fig:setup}(c, d)]. Three other dies had no slots written on them. 
The resonators were then patterned photolithographically and aligned to the previously etched slot patterns, followed by a second RIE step that transferred the resonator pattern into the Al film. 
For the dies with slots, these follow the entire length of each resonator up to the feedline coupling elbow. A
slight misalignment during the photolithography step caused all of the resonators to be shifted by 1.5 $\mu$m so that the slots are offset from the centerline of each resonator by this amount. 

We cool the resonators on a ${\rm ^3}$He refrigerator and we measure the complex transmission $S_{21}$ through the feedline using a 
a vector network analyzer (Agilent N5230A). 
We employ a series of cold attenuators on the input side of the feedline and a cryogenic HEMT amplifier (gain $\approx 38$~dB between $0.5 - 11$~GHz; $T_N \approx 5$~K) on the output side.
A power of about -90~dBm delivered to the feedline keeps the resonators comfortably in the linear regime. 
We use a superconducting Helmholtz coil to generate a magnetic field and a $\mu$-metal cylinder attenuates stray fields in the laboratory. 
For each value of $B$, we heat the sample above $T_c $ to $1.4$~K, adjust the current through our Helmholtz coil to the desired value, then cool down to 310 $\pm$ 0.2~mK. 

\begin{figure}
\centering
\includegraphics[width=3.35in]{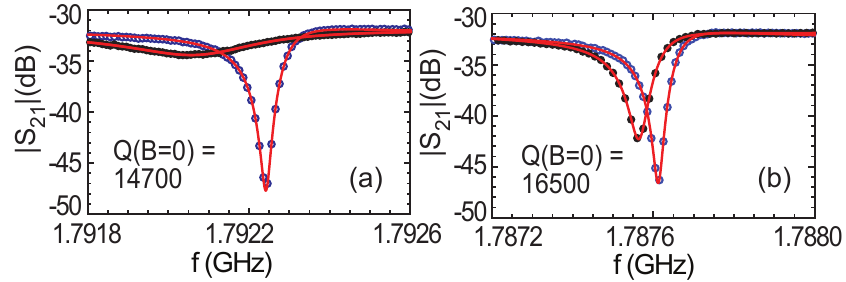}
  \caption{(Color online) $|S_{21}|(f)$ at $B=$ 0 (blue, open circles) and $B=$ 86~$\mu$T (black, closed circles) for resonator with (a) no slot;
  (b) $200$~nm-wide slot. 
  Lines correspond to fits as described in text. 
  \label{fig:dips-slot&no_slot}}
\end{figure}

In general, the addition of vortices through field-cooling results in a downward shift in the resonance frequency and a reduction in the quality factor. 
In Figure \ref{fig:dips-slot&no_slot} we plot $|S_{21}|(f)$ for the resonators near 1.8~GHz, from a chip with no slot [Fig. \ref{fig:dips-slot&no_slot}(a)] and from a chip with a slot [Fig. \ref{fig:dips-slot&no_slot}(b)], where both resonators were measured at $B=$ 0 and $B=$ 86~$\mu$T. 
For $B=$ 0, thus, with no vortices present, both resonators have comparable quality factors, determined primarily by the coupling to the feedline. 
However, the resonators behave quite differently when cooled in the same magnetic field.  
For the chip with no slot, this results in a substantial broadening and downward frequency shift. In contrast, the resonator with the slot has only a minimal reduction in its resonance linewidth and a small frequency shift.

We can make a quantitative comparison between the resonators with and without slots by fitting the resonance trajectories in the complex plane to extract the quality factor $Q$ and resonance frequency $f_0$ for each resonator at the various cooling fields. 
The fit process was the same as in Ref. \cite{song09}, which followed a similar $10$-parameter fitting procedure to what has been done for MKID measurements \cite{mazin04}. 
We define the excess loss in each resonator due to the presence of vortices, $1/Q_v(B) = 1/Q(B) - 1/Q(B=0)$, 
thus removing the loss due to thermal quasiparticles, coupling to the feedline, and any other field-independent loss mechanisms. 
In a similar manner, we compute the fractional frequency shift of each resonance, $\delta f/f_0(B) =\left[f_0(B=0)-f_0(B)\right]/f_0(B=0)$. 

\begin{figure}
\centering
\includegraphics[width=3.35in]{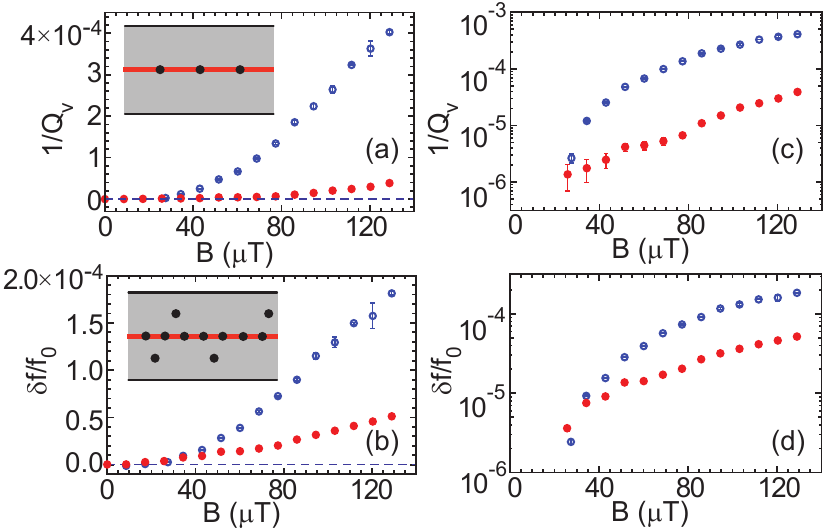}
\caption{(Color online) Comparison of resonators without (blue, open circles) and with a slot (red, closed circles): (a) $1/Q_v (B)$; (b) $\delta f/f_0 (B)$. 
(c, d) Data from (a, b) plotted on a log scale. Insets show predicted vortex arrangements with a slot for small $B$ (upper) and $B>$ 80 $\mu$T (lower).  
\label{fig:loss&shift-all}}
\end{figure}

In Figure \ref{fig:loss&shift-all} we plot $1/Q_v (B)$ and $\delta f/f_0 (B)$ for the resonators near 1.8~GHz with and without a slot. 
Error bars from the fitting process are included, but in most cases are too small to be seen.  
Above a certain onset cooling field of $\sim$25 $\mu$T, both $1/Q_v$ and $\delta f/f_0$ increase with $B$ for both resonators, however, for the resonator with a slot, these quantities are much lower than in the resonator without a slot. Viewing the data on a logarithmic scale emphasizes 
the effect of the slot: a reduction in $1/Q_v$ by a factor between 10-20 over almost the entire measured field range relative to the resonator without a slot [Fig. \ref{fig:loss&shift-all}(c)]. 
The slot also reduces $\delta f/f_0(B)$, although by a smaller factor [Fig. \ref{fig:loss&shift-all}(d)]. 

We note that a comparison of the data in Figure \ref{fig:loss&shift-all} for the resonator with no slot with our earlier measurements in Al films from Ref.~\cite{song09} indicates a smaller $1/Q_v$ in the more recent resonators without slots by a factor of $\sim$~0.5. While both resonators had the same layout with a similar fabrication process, the Al film from Ref.~\cite{song09} had a RRR of 10, while the Al film in the resonators described in this work had a RRR of 19. This turns out to be the two extremes of RRR that we have observed from a series of deposited Al films on sapphire in our evaporator. It appears that 
the resulting variation in the 
mean free path 
affects the loss contributed by a vortex. We are currently investigating this effect in more detail. 
Nonetheless, we emphasize that all of the resonators presented in this work, both with and without slots, were fabricated simultaneously from a single Al film. Furthermore, in separate cooldowns we have measured two more chips from this wafer, one with a slot and one without, and obtained quite similar results to Figs. 2, 3.  
Thus, the differences in vortex response that we observe here are due solely to the presence of the slot in the resonator. 

Vortex distributions in superconducting strips cooled in perpendicular magnetic fields have been studied theoretically \cite{likharev72, clem98, maksimova98, bronson06} using energetic considerations of the interaction between the vortices and screening currents in the strip. This leads to a prediction of a threshold cooling field, 
below which the strip remains flux free. Beyond the threshold field, the vortex density in the strip increases with $B$, with vortices initially trapped in a single row along the strip centerline. For larger $B$, the vortex distribution evolves into progressively more parallel rows \cite{bronson06}. 
These predictions have been confirmed experimentally through vortex imaging \cite{stan04, kuit08}, and are also consistent with microwave resonator measurements \cite{song09}. 
Thus, for our resonators, both with and without a slot, we expect the initial trapped vortices to be located near the centerline of the resonator. On the resonators with a slot, these initial vortices are almost certainly located in the slot [Fig. \ref{fig:loss&shift-all}(a) inset], where the vortex line energy is reduced, despite the slight misalignment of the slot. 
At some threshold field, when the vortex density in the slot becomes too large, vortices will begin to get trapped outside of the slot [Fig. \ref{fig:loss&shift-all}(b) inset], and thus contribute greater loss because of the weaker pinning there. In Fig. \ref{fig:loss&shift-all}(c), we observe a kink in $1/Q_v$ near 80 $\mu$T for the resonator with a slot, above which the loss increases more rapidly. This kink, which would correspond to a vortex spacing of about 2 $\mu$m for a single row of vortices in the slot, may indicate such a threshold. 
Further reduction of $1/Q_v$ at larger $B$ may be possible by patterning multiple parallel slots. 

Although we have focused our comparison here on the resonators near 1.8 GHz, we have observed reductions in $1/Q_v$ between the higher frequency resonators with and without slots by factors of $\sim$8, 6, and 3 for the resonators near 3.3, 6.9, and 11.0 GHz, respectively. For a particular level of enhanced pinning, the loss will be larger as the frequency increases towards $f_d$. With no slot, $1/Q_v$ can actually decrease with frequency \cite{song09}, such that the 
reduction of $1/Q_v$ for a resonator with a slot compared to one with no slot will be less substantial at higher frequencies. 

Based on our measurements here and in Ref.~\cite{song09}, the presence of even a small number of vortices has a significant influence on the resonator quality factor. 
A single narrow slot along the centerline of an Al CPW resonator provides a straightforward method to increase the pinning and reduce the loss from vortices by over an order of magnitude.

We acknowledge useful discussions with R. McDermott and J.M. Martinis. This work was supported by the National Science Foundation under Grant DMR-0547147. Device fabrication was performed at the Cornell NanoScale Facility.


\begin{thebibliography}{0}
\expandafter\ifx\csname natexlab\endcsname\relax\def\natexlab#1{#1}\fi
\expandafter\ifx\csname bibnamefont\endcsname\relax
  \def\bibnamefont#1{#1}\fi
\expandafter\ifx\csname bibfnamefont\endcsname\relax
  \def\bibfnamefont#1{#1}\fi
\expandafter\ifx\csname citenamefont\endcsname\relax
  \def\citenamefont#1{#1}\fi
\expandafter\ifx\csname url\endcsname\relax
  \def\url#1{\texttt{#1}}\fi
\expandafter\ifx\csname urlprefix\endcsname\relax\def\urlprefix{URL }\fi
\providecommand{\bibinfo}[2]{#2}
\providecommand{\eprint}[2][]{\url{#2}}

\end{thebibliography}


\begin{thebibliography}{22}
\expandafter\ifx\csname natexlab\endcsname\relax\def\natexlab#1{#1}\fi
\expandafter\ifx\csname bibnamefont\endcsname\relax
  \def\bibnamefont#1{#1}\fi
\expandafter\ifx\csname bibfnamefont\endcsname\relax
  \def\bibfnamefont#1{#1}\fi
\expandafter\ifx\csname citenamefont\endcsname\relax
  \def\citenamefont#1{#1}\fi
\expandafter\ifx\csname url\endcsname\relax
  \def\url#1{\texttt{#1}}\fi
\expandafter\ifx\csname urlprefix\endcsname\relax\def\urlprefix{URL }\fi
\providecommand{\bibinfo}[2]{#2}
\providecommand{\eprint}[2][]{\url{#2}}

\bibitem[{\citenamefont{Clarke and Wilhelm}(2008)}]{clarke08}
\bibinfo{author}{\bibfnamefont{J.}~\bibnamefont{Clarke}} \bibnamefont{and}
  \bibinfo{author}{\bibfnamefont{F.~K.} \bibnamefont{Wilhelm}},
  \bibinfo{journal}{Nature} \textbf{\bibinfo{volume}{453}},
  \bibinfo{pages}{1031} (\bibinfo{year}{2008}).

\bibitem[{\citenamefont{Wallraff et~al.}(2004)\citenamefont{Wallraff, Schuster,
  Blais, Frunzio, Huang, Majer, Kumar, Girvin, and Schoelkopf}}]{wallraff04}
\bibinfo{author}{\bibfnamefont{A.}~\bibnamefont{Wallraff}},
  \bibinfo{author}{\bibfnamefont{D.}~\bibnamefont{Schuster}},
  \bibinfo{author}{\bibfnamefont{A.}~\bibnamefont{Blais}},
  \bibinfo{author}{\bibfnamefont{L.}~\bibnamefont{Frunzio}},
  \bibinfo{author}{\bibfnamefont{R.}~\bibnamefont{Huang}},
  \bibinfo{author}{\bibfnamefont{J.}~\bibnamefont{Majer}},
  \bibinfo{author}{\bibfnamefont{S.}~\bibnamefont{Kumar}},
  \bibinfo{author}{\bibfnamefont{S.}~\bibnamefont{Girvin}}, \bibnamefont{and}
  \bibinfo{author}{\bibfnamefont{R.}~\bibnamefont{Schoelkopf}},
  \bibinfo{journal}{Nature} \textbf{\bibinfo{volume}{431}},
  \bibinfo{pages}{162} (\bibinfo{year}{2004}).

\bibitem[{\citenamefont{Hofheinz et~al.}(2009)\citenamefont{Hofheinz, Wang,
  Ansmann, Bialczak, Lucero, Neeley, O'Connell, Sank, Wenner, Martinis
  et~al.}}]{hofheinz09}
\bibinfo{author}{\bibfnamefont{M.}~\bibnamefont{Hofheinz}},
  \bibinfo{author}{\bibfnamefont{H.}~\bibnamefont{Wang}},
  \bibinfo{author}{\bibfnamefont{M.}~\bibnamefont{Ansmann}},
  \bibinfo{author}{\bibfnamefont{R.}~\bibnamefont{Bialczak}},
  \bibinfo{author}{\bibfnamefont{E.}~\bibnamefont{Lucero}},
  \bibinfo{author}{\bibfnamefont{M.}~\bibnamefont{Neeley}},
  \bibinfo{author}{\bibfnamefont{A.}~\bibnamefont{O'Connell}},
  \bibinfo{author}{\bibfnamefont{D.}~\bibnamefont{Sank}},
  \bibinfo{author}{\bibfnamefont{J.}~\bibnamefont{Wenner}},
  \bibinfo{author}{\bibfnamefont{J.}~\bibnamefont{Martinis}},
  \bibnamefont{et~al.}, \bibinfo{journal}{Nature}
  \textbf{\bibinfo{volume}{459}}, \bibinfo{pages}{546} (\bibinfo{year}{2009}).

\bibitem[{\citenamefont{Zmuidzinas et~al.}(2003)\citenamefont{Zmuidzinas,
  Vayonakis, Day, LeDuc, and Mazin}}]{zmuidzinas03}
\bibinfo{author}{\bibfnamefont{J.}~\bibnamefont{Zmuidzinas}},
  \bibinfo{author}{\bibfnamefont{A.}~\bibnamefont{Vayonakis}},
  \bibinfo{author}{\bibfnamefont{P.~K.} \bibnamefont{Day}},
  \bibinfo{author}{\bibfnamefont{H.~G.} \bibnamefont{LeDuc}}, \bibnamefont{and}
  \bibinfo{author}{\bibfnamefont{B.~A.} \bibnamefont{Mazin}},
  \bibinfo{journal}{Nature} \textbf{\bibinfo{volume}{425}},
  \bibinfo{pages}{817} (\bibinfo{year}{2003}).

\bibitem[{\citenamefont{O'Connell et~al.}(2008)\citenamefont{O'Connell,
  Ansmann, Bialczak, Hofheinz, Katz, Lucero, McKenney, Neeley, Wang, Weig
  et~al.}}]{oconnell08}
\bibinfo{author}{\bibfnamefont{A.}~\bibnamefont{O'Connell}},
  \bibinfo{author}{\bibfnamefont{M.}~\bibnamefont{Ansmann}},
  \bibinfo{author}{\bibfnamefont{R.}~\bibnamefont{Bialczak}},
  \bibinfo{author}{\bibfnamefont{M.}~\bibnamefont{Hofheinz}},
  \bibinfo{author}{\bibfnamefont{N.}~\bibnamefont{Katz}},
  \bibinfo{author}{\bibfnamefont{E.}~\bibnamefont{Lucero}},
  \bibinfo{author}{\bibfnamefont{C.}~\bibnamefont{McKenney}},
  \bibinfo{author}{\bibfnamefont{M.}~\bibnamefont{Neeley}},
  \bibinfo{author}{\bibfnamefont{H.}~\bibnamefont{Wang}},
  \bibinfo{author}{\bibfnamefont{E.}~\bibnamefont{Weig}}, \bibnamefont{et~al.},
  \bibinfo{journal}{Appl. Phys. Lett.} \textbf{\bibinfo{volume}{92}},
  \bibinfo{pages}{112903} (\bibinfo{year}{2008}).

\bibitem[{\citenamefont{Song et~al.}(2009)\citenamefont{Song, Heitmann, DeFeo,
  Yu, McDermott, Neeley, Martinis, and Plourde}}]{song09}
\bibinfo{author}{\bibfnamefont{C.}~\bibnamefont{Song}},
  \bibinfo{author}{\bibfnamefont{T.~W.} \bibnamefont{Heitmann}},
  \bibinfo{author}{\bibfnamefont{M.~P.} \bibnamefont{DeFeo}},
  \bibinfo{author}{\bibfnamefont{K.}~\bibnamefont{Yu}},
  \bibinfo{author}{\bibfnamefont{R.}~\bibnamefont{McDermott}},
  \bibinfo{author}{\bibfnamefont{M.}~\bibnamefont{Neeley}},
  \bibinfo{author}{\bibfnamefont{J.~M.} \bibnamefont{Martinis}},
  \bibnamefont{and} \bibinfo{author}{\bibfnamefont{B.~L.~T.}
  \bibnamefont{Plourde}}, \bibinfo{journal}{Phys. Rev. B}
  \textbf{\bibinfo{volume}{79}}, \bibinfo{pages}{174512}
  (\bibinfo{year}{2009}).

\bibitem[{\citenamefont{Coffey and Clem}(1991)}]{coffey91}
\bibinfo{author}{\bibfnamefont{M.~W.} \bibnamefont{Coffey}} \bibnamefont{and}
  \bibinfo{author}{\bibfnamefont{J.~R.} \bibnamefont{Clem}},
  \bibinfo{journal}{Phys. Rev. Lett.} \textbf{\bibinfo{volume}{67}},
  \bibinfo{pages}{386} (\bibinfo{year}{1991}).

\bibitem[{\citenamefont{Brandt}(1991)}]{brandt91}
\bibinfo{author}{\bibfnamefont{E.~H.} \bibnamefont{Brandt}},
  \bibinfo{journal}{Phys. Rev. Lett.} \textbf{\bibinfo{volume}{67}},
  \bibinfo{pages}{2219} (\bibinfo{year}{1991}).

\bibitem[{\citenamefont{Gittleman and Rosenblum}(1968)}]{gittleman68}
\bibinfo{author}{\bibfnamefont{J.~I.} \bibnamefont{Gittleman}}
  \bibnamefont{and}
  \bibinfo{author}{\bibfnamefont{B.}~\bibnamefont{Rosenblum}},
  \bibinfo{journal}{J. Appl. Phys.} \textbf{\bibinfo{volume}{39}},
  \bibinfo{pages}{2617} (\bibinfo{year}{1968}).

\bibitem[{\citenamefont{Pompeo and Silva}(2008)}]{pompeo08}
\bibinfo{author}{\bibfnamefont{N.}~\bibnamefont{Pompeo}} \bibnamefont{and}
  \bibinfo{author}{\bibfnamefont{E.}~\bibnamefont{Silva}},
  \bibinfo{journal}{Phys. Rev. B} \textbf{\bibinfo{volume}{78}},
  \bibinfo{pages}{094503} (\bibinfo{year}{2008}).

\bibitem[{\citenamefont{Zaitsev et~al.}(2007)\citenamefont{Zaitsev, Schneider,
  Hott, Schwarz, and Geerk}}]{zaitsev07}
\bibinfo{author}{\bibfnamefont{A.~G.} \bibnamefont{Zaitsev}},
  \bibinfo{author}{\bibfnamefont{R.}~\bibnamefont{Schneider}},
  \bibinfo{author}{\bibfnamefont{R.}~\bibnamefont{Hott}},
  \bibinfo{author}{\bibfnamefont{T.}~\bibnamefont{Schwarz}}, \bibnamefont{and}
  \bibinfo{author}{\bibfnamefont{J.}~\bibnamefont{Geerk}},
  \bibinfo{journal}{Phys. Rev. B} \textbf{\bibinfo{volume}{75}},
  \bibinfo{pages}{212505} (\bibinfo{year}{2007}).

\bibitem[{\citenamefont{Belk et~al.}(1996)\citenamefont{Belk, Oates, Feld,
  Dresselhaus, and Dresselhaus}}]{belk96}
\bibinfo{author}{\bibfnamefont{N.}~\bibnamefont{Belk}},
  \bibinfo{author}{\bibfnamefont{D.~E.} \bibnamefont{Oates}},
  \bibinfo{author}{\bibfnamefont{D.~A.} \bibnamefont{Feld}},
  \bibinfo{author}{\bibfnamefont{G.}~\bibnamefont{Dresselhaus}},
  \bibnamefont{and} \bibinfo{author}{\bibfnamefont{M.~S.}
  \bibnamefont{Dresselhaus}}, \bibinfo{journal}{Phys. Rev. B}
  \textbf{\bibinfo{volume}{53}}, \bibinfo{pages}{3459} (\bibinfo{year}{1996}).

\bibitem[{\citenamefont{Daldini et~al.}(1974)\citenamefont{Daldini, Martinoli,
  Olsen, and Berner}}]{daldini74}
\bibinfo{author}{\bibfnamefont{O.}~\bibnamefont{Daldini}},
  \bibinfo{author}{\bibfnamefont{P.}~\bibnamefont{Martinoli}},
  \bibinfo{author}{\bibfnamefont{J.~L.} \bibnamefont{Olsen}}, \bibnamefont{and}
  \bibinfo{author}{\bibfnamefont{G.}~\bibnamefont{Berner}},
  \bibinfo{journal}{Phys. Rev. Lett.} \textbf{\bibinfo{volume}{32}},
  \bibinfo{pages}{218} (\bibinfo{year}{1974}).

\bibitem[{\citenamefont{Plourde et~al.}(2002)\citenamefont{Plourde,
  Van~Harlingen, Saha, Besseling, Hesselberth, and Kes}}]{plourde02}
\bibinfo{author}{\bibfnamefont{B.~L.~T.} \bibnamefont{Plourde}},
  \bibinfo{author}{\bibfnamefont{D.~J.} \bibnamefont{Van~Harlingen}},
  \bibinfo{author}{\bibfnamefont{N.}~\bibnamefont{Saha}},
  \bibinfo{author}{\bibfnamefont{R.}~\bibnamefont{Besseling}},
  \bibinfo{author}{\bibfnamefont{M.~B.~S.} \bibnamefont{Hesselberth}},
  \bibnamefont{and} \bibinfo{author}{\bibfnamefont{P.~H.} \bibnamefont{Kes}},
  \bibinfo{journal}{Phys. Rev. B} \textbf{\bibinfo{volume}{66}},
  \bibinfo{pages}{054529} (\bibinfo{year}{2002}).

\bibitem[{\citenamefont{Sonnet}(2005)}]{sonnet05}
\bibinfo{author}{\bibnamefont{Sonnet}}, \emph{\bibinfo{title}{{}}}
  (\bibinfo{publisher}{Sonnet Software, Inc., Liverpool, NY},
  \bibinfo{year}{2005}).

\bibitem[{\citenamefont{Mazin}(2004)}]{mazin04}
\bibinfo{author}{\bibfnamefont{B.~A.} \bibnamefont{Mazin}}, Ph.D. thesis,
  \bibinfo{school}{Caltech} (\bibinfo{year}{2004}).

\bibitem[{\citenamefont{Likharev}(1971)}]{likharev72}
\bibinfo{author}{\bibfnamefont{K.~K.} \bibnamefont{Likharev}},
  \bibinfo{journal}{Sov. Radiophys.} \textbf{\bibinfo{volume}{14}},
  \bibinfo{pages}{722} (\bibinfo{year}{1971}).

\bibitem[{\citenamefont{Clem}(1998)}]{clem98}
\bibinfo{author}{\bibfnamefont{J.~R.} \bibnamefont{Clem}},
  \bibinfo{journal}{Bull. Am. Phys. Soc.} \textbf{\bibinfo{volume}{43}},
  \bibinfo{pages}{401} (\bibinfo{year}{1998}).

\bibitem[{\citenamefont{Maksimova}(1998)}]{maksimova98}
\bibinfo{author}{\bibfnamefont{G.~M.} \bibnamefont{Maksimova}},
  \bibinfo{journal}{Phys. Solid State} \textbf{\bibinfo{volume}{40}},
  \bibinfo{pages}{1607} (\bibinfo{year}{1998}).

\bibitem[{\citenamefont{Bronson et~al.}(2006)\citenamefont{Bronson, Gelfand,
  and Field}}]{bronson06}
\bibinfo{author}{\bibfnamefont{E.}~\bibnamefont{Bronson}},
  \bibinfo{author}{\bibfnamefont{M.~P.} \bibnamefont{Gelfand}},
  \bibnamefont{and} \bibinfo{author}{\bibfnamefont{S.~B.} \bibnamefont{Field}},
  \bibinfo{journal}{Phys. Rev. B} \textbf{\bibinfo{volume}{73}},
  \bibinfo{pages}{144501} (\bibinfo{year}{2006}).

\bibitem[{\citenamefont{Stan et~al.}(2004)\citenamefont{Stan, Field, and
  Martinis}}]{stan04}
\bibinfo{author}{\bibfnamefont{G.}~\bibnamefont{Stan}},
  \bibinfo{author}{\bibfnamefont{S.~B.} \bibnamefont{Field}}, \bibnamefont{and}
  \bibinfo{author}{\bibfnamefont{J.~M.} \bibnamefont{Martinis}},
  \bibinfo{journal}{Phys. Rev. Lett.} \textbf{\bibinfo{volume}{92}},
  \bibinfo{pages}{97003} (\bibinfo{year}{2004}).

\bibitem[{\citenamefont{Kuit et~al.}(2008)\citenamefont{Kuit, Kirtley, van~der
  Veur, Molenaar, Roesthuis, Troeman, Clem, Hilgenkamp, Rogalla, and
  Flokstra}}]{kuit08}
\bibinfo{author}{\bibfnamefont{K.~H.} \bibnamefont{Kuit}},
  \bibinfo{author}{\bibfnamefont{J.~R.} \bibnamefont{Kirtley}},
  \bibinfo{author}{\bibfnamefont{W.}~\bibnamefont{van~der Veur}},
  \bibinfo{author}{\bibfnamefont{C.~G.} \bibnamefont{Molenaar}},
  \bibinfo{author}{\bibfnamefont{F.~J.~G.} \bibnamefont{Roesthuis}},
  \bibinfo{author}{\bibfnamefont{A.~G.~P.} \bibnamefont{Troeman}},
  \bibinfo{author}{\bibfnamefont{J.~R.} \bibnamefont{Clem}},
  \bibinfo{author}{\bibfnamefont{H.}~\bibnamefont{Hilgenkamp}},
  \bibinfo{author}{\bibfnamefont{H.}~\bibnamefont{Rogalla}}, \bibnamefont{and}
  \bibinfo{author}{\bibfnamefont{J.}~\bibnamefont{Flokstra}},
  \bibinfo{journal}{Phys. Rev. B} \textbf{\bibinfo{volume}{77}},
  \bibinfo{pages}{134504} (\bibinfo{year}{2008}).

\end{thebibliography}

\end{document}